\documentclass{cargese}\usepackage{graphicx}
\let\footnote\savefootnote
\let\footnotetext\savefootnotetext 
\let\footnoterule\savefootnoterule 
\normallatexbib

\def\beq{\begin{equation}}
\def\eeq{\end{equation}}
\def\beqa{\begin{eqnarray}}
\def\eeqa{\end{eqnarray}}
\renewcommand{\a}{\alpha}

\newcommand{\eqn}[1]{(\ref{#1})}

\def\NS{{\rm NS}}
\def\R{{\rm R}}
\begin{document}


\articletitle[Type 0 D-branes]{Anomalous couplings of type 0 \\ D-branes}


\author{Marco Bill\'o}


\affil{Dipartimento di Fisica Teorica, Universit\`a di
Torino and\\
I.N.F.N., Sezione di Torino, via P. Giuria 1, I-10125, Torino, Italy }         

%


%

\author{Ben Craps and Frederik Roose }
\affil{Instituut voor Theoretische Fysica\\
Katholieke Universiteit Leuven, B-3001 Leuven, Belgium }

\begin{abstract}
Closed type 0 string theories and their D-branes are introduced. The full
Wess-Zumino action of these D-branes is derived. The analogy with type II is
emphasized throughout the argument.
\end{abstract}

\paragraph{Type 0 Strings and D-branes}
\index{type 0 D-branes}
Type 0 string theory  is a non-supersymmetric, modular invariant 
theory  of
closed strings. The presence of a tachyonic state in their spectrum makes type 0
strings much harder to analyse than the supersymmetric type II strings.
Nevertheless, type 0 and type II strings have many features in common. 
In this contribution this fact will be exploited to show that type 0 D-branes
have anomalous terms in their worldvolume action. The analysis is based on
Ref.~\cite{BCR}, to which we refer for a more detailed treatment and for a more
complete list of references.

In the Neveu-Schwarz-Ramond formulation, type II string theories are obtained 
by imposing  independent GSO projections on the left and right moving part. 
This amounts to keeping the following (left,right) sectors: 
\begin{eqnarray}
{\rm IIB:}&&(\NS+,\NS+)\ ,~~(\R+,\R+)\ ,~~(\R+,\NS+)\ ,~~(\NS+,\R+)~;\nonumber\\
{\rm IIA:}&&(\NS+,\NS+)\ ,~~(\R+,\R-)\ ,~~(\R+,\NS+)\ ,~~(\NS+,\R-)~,\nonumber
\end{eqnarray}
where for instance R$\pm$ is the Ramond sector projected with 
$P_{\rm GSO}=(1 \pm (-)^F)/2$,
$F$ being the world-sheet fermion number.

The type 0 string theories contain instead the following sectors:
\begin{eqnarray}
{\rm 0B:}&&(\NS+,\NS+)\ ,~~(\NS-,\NS-)\ ,~~(\R+,\R+)\ ,~~(\R-,\R-)~; \nonumber\\
{\rm 0A:}&&(\NS+,\NS+)\ ,~~(\NS-,\NS-)\ ,~~(\R+,\R-)\ ,~~(\R-,\R+)~.\nonumber
\end{eqnarray}
These theories do not contain bulk spacetime fermions, which would have to
come from ``mixed'' (R,NS) sectors. 
The inclusion of the NS-NS sectors with odd fermion numbers means that the 
closed string tachyon is not projected out. 
The third difference with type II theories is that the R-R spectrum is 
doubled.
For instance, in the 0B case, beside the IIB R-R potentials $C_{p+1}$
contained in
bispinors of the (R+,R+) sector, there are the potentials $C'_{p+1}$ from 
bispinors of the (R$-$,R$-$) sector.  
Note that the bispinors containing the primed and unprimed R-R potentials
have  opposite chirality.  This implies a sign difference in the
Poincar\'e duality relations among the field strengths. 
Thus, for instance, in type 0B there is an unconstrained  five-form
field strength, whose self-dual (anti-self-dual) part is the unprimed (primed)
field strength.

For our purposes, convenient combinations of $C_{p+1}$ and $C'_{p+1}$ are
\beq \label{C+-}
(C_{p+1})_\pm=\frac{1}{\sqrt{2}}(C_{p+1}\pm C'_{p+1})~.
\eeq 
For $p=3$ these are the electric ($+$) and magnetic ($-$) potentials 
\cite{BG}. We will adopt this terminology also for other values of $p$. 
There turn out to be 
four types of ``elementary'' D-branes for each $p$: an electric and a 
magnetic one ({\it i.e.}, charged under $(C_{p+1})_\pm$), and the
corresponding antibranes \cite{BG}.  
\paragraph{Anomalous Couplings}
\index{anomalous couplings}
The open strings stretching between two like branes are bosons, just like the 
bulk fields of type 0. However, a boundary state computation shows that
fermions appear from strings between an 
electric and a magnetic brane \cite{BG}. Thus one could
wonder whether there are chiral fermions on the intersection of an electric 
and a magnetic brane. Consider such an orthogonal intersection with no overall 
transverse directions. If the dimension of the intersection is
two or six, the computation reveals that there are precisely enough 
fermionic degrees of freedom on the intersection 
to form one chiral fermion.

In type II string theory, the analogous computation shows that chiral fermions 
are present on two or six dimensional intersections of two orthogonal branes 
with no overall transverse directions. That observation has
had far reaching consequences. Namely, the presence of chiral fermions has been shown to
lead to gauge and gravitational anomalies on those intersections of D-branes 
\cite{GHM}. In a consistent theory, such anomalies should be cancelled by 
anomaly inflow\index{anomaly inflow}. In the
present case, the anomaly inflow is  provided by the anomalous D-brane couplings in the Wess-Zumino part of
the D-brane action \cite{GHM}. These anomalous couplings have an anomalous variation localized on the
intersections with other branes.     

To sketch how this anomaly inflow comes about, let us focus on the case of two 
type IIB D5-branes (to be denoted by D5 and D5') intersecting on a string. 
The Wess-Zumino action on  D5 contains a term of the form
$\int_{\rm D5}C_2\wedge Y_4$, where $C_2$ is the R-R two-form potential 
and $Y_4$ a certain four-form involving the field strength of gauge field on 
D5 and the curvature two-forms of the tangent and normal bundles of D5. To be precise, one
should replace this term by $\int_{\rm D5}H_3\wedge \omega_3$, with 
$Y_4=d\omega_3$ and $H_3$ the complete
gauge-invariant field strength of $C_2$ (which generically differs from $dC_2$).
Since the gauge variation of the ``Chern-Simons''
form $\omega_3$ is given by $\delta\omega_3=dI_2$ for some two-form $I_2$,
the anomalous term on D5 have a variation localized on the intersection 
with D5':
\beq	
\delta \int_{{\rm D}5}H_3\wedge \omega_3=\int_{{\rm D}5}dH_3\wedge I_2=
\int_{{\rm D}5}d*H_7 \wedge I_2=
\int_{{\rm D}5}\delta_{{\rm D}5'}\wedge I_2~,
\eeq 
which can thus cancel the anomaly due to the chiral fermions living on the
intersection.  
A careful analysis of all the anomalies \cite{GHM} shows that the anomalous 
part of the D$p$-brane action\index{Wess-Zumino action} is given, in terms of the formal sum $C$
of the various R-R forms, by
\beq
S_{\rm WZ}=\frac{T_p}{\kappa}\int_{p+1} C\wedge e^{2\pi\a '\,F+B}\wedge
\sqrt{\hat{A}(R_T)/\hat{A}(R_N)}~~.
\eeq
Here $T_p/\kappa$ denotes the D$p$-brane tension, $F$ the gauge field on 
the brane and $B$ the NS-NS two-form.
Further, 
$R_T$ and $R_N$ are the curvatures of the tangent and normal 
bundles of the D-brane world-volume, and $\hat{A}$ denotes the A-roof genus.

Let us now return to type 0 string theory. As stated above, here chiral fermions live on intersections of 
electric and magnetic type 0 D-branes.  
The associated gauge and gravitational anomalies on such intersections 
match the ones for type II D-branes. To cancel them, the minimal coupling of a
D$p$-brane to a ($p+1$)-form R-R potential should be extended to the following 
Wess-Zumino action \cite{BCR}:   
\beq
\label{WZaction}
S_{\rm WZ}=\frac{T_p}{\kappa}\int_{p+1}(C)_\pm\wedge 
{\rm e}^{2\pi\a '\,F+B}\wedge
\sqrt{\hat{A}(R_T)/\hat{A}(R_N)}~.
\eeq
The $\pm$ in Eq. \eqn{WZaction} distinguishes between electric and magnetic 
branes. Note that $T_p/\kappa$ denotes the tension
of a type II D$p$-brane, which can be computed to be $\sqrt 2$ times the 
type 0 D$p$-brane tension \cite{BG}.

The argument that the variation of this action
cancels the anomaly on the intersection is a copy of
the one described above in the type II case, apart from one slight
subtlety. For definiteness, consider the intersection of an electric and a 
magnetic D$5$-brane on a string.
Varying the electric D$5$-brane action (exhibiting the $(C_2)_+$ potential, 
or rather, its field strength $(H_3)_+$), one finds that the variation 
is localized on the intersection of the electric  D$5$-brane with
branes charged magnetically under the $(H_3)_+$ field strength. 
Using Eq. (\ref{C+-}), the different behaviour under Poincar\'e duality 
of the primed and unprimed R-R field strengths 
 shows that these are precisely the branes  
carrying (electric) $(H_7)_-$ charge,  i.e. 
what we called the magnetic D$5$-branes. 
Schematically,
\beq
\delta \int_{{\rm D}5_+}(H_3)_+\wedge \omega_3=\int_{{\rm D}5_+}d(H_3)_+
\wedge I_2= \int_{{\rm D}5_+}d*(H_7)_-\wedge I_2=
\int_{{\rm D}5_+}\delta_{{\rm D}5_-}\wedge I_2~.
\eeq
A completely analogous discussion goes through for the
variation of the magnetic D$5$-brane action.


\begin{acknowledgments}
B.C. and F.R. would like to thank the organisers for a very nice school, and for
financial support. This work was supported by the European Commission TMR 
programme ERBFMRX-CT96-0045. B.C. is Aspirant FWO-Vlaanderen.
\end{acknowledgments}



%

\end{document}